\documentclass[aps,prl,twocolumn,showpacs]{revtex4}
\usepackage{epsf}
\usepackage{amsmath,amssymb}
\usepackage{graphicx}
\usepackage[english]{babel}
\usepackage{color}

\begin{document}

\title{Quantum Dynamics Against a Noisy Background}

\date{\today}

\author{Valentin V. Sokolov}
\affiliation{Budker Institute of Nuclear Physics, Novosibirsk,
Russia}
\affiliation{Novosibirsk Technical University}
\author{Oleg V. Zhirov}
\affiliation{Budker Institute of Nuclear Physics, Novosibirsk,
Russia}
\affiliation{Novosibirsk State University}
\author{Yaroslav A. Kharkov}
\affiliation{Novosibirsk State University}
\affiliation{Budker Institute of Nuclear Physics, Novosibirsk,
Russia}

\begin{abstract} 
By the example of a kicked quartic oscillator we
investigate the dynamics of classically chaotic quantum systems
with few degrees of freedom affected by persistent external
noise. Stability and reversibility of the motion are analyzed
in detail in dependence on the noise level
$\sigma$. The critical level $\sigma_c(t)$, below which the
response of the system to the noise remains weak, is studied
versus the evolution time. In the regime with the Ehrenfest
time interval $t_E$ so short that the classical Lyapunov
exponential decay of the Peres fidelity does not show up the
time dependence of this critical value is proved to be
power-like. We estimate also the decoherence time after which
the motion turns into a Markovian process.
\end{abstract}


\maketitle

Exponential sensitivity of nonlinear classical systems which
display chaotic behavior to arbitrarily weak perturbations makes
impractical the treatment of such systems as closed ones.
Inevitably, the interaction with environment crucially influences
the dynamics of such systems. In many cases this influence can
be considered as a noise, that turns the motion into an
irreversible random process.

Quite opposite, the quantum dynamics of the same systems
manifests a considerable degree of stability against
external perturbations \cite{arrow1}. A quantitative analysis
shows \cite{arrow2} that the sensitivity to an instant perturbation
is of a threshold nature: there exists a critical value $\xi_c$ of
the strength $\xi$ of the perturbation, below which the response
of the system remains weak. This critical value depends on complexity
of the quantum Wigner function, that can be characterized, for example,
by the number of its Fourier harmonics.

In general, the critical strength
$\sigma_c(t)$ decreases exponentially 
within the Ehrenfest interval $0<t<t_E$ during which the Wigner
function still satisfies the classical Liouville equation and
therefore the role of the noisy environment remains decisive.
This fact is very important \cite{Kolovsky94} for understanding
the quantum-classical correspondence in the nontrivial case of
classically chaotic systems. However, in the case
when the Ehrenfest interval is so short that the classical
exponential instability has not enough time to show up,
in-depth information on chaotic quantum dynamics against a
noisy background is quite limited up to now.

In this paper we present an advanced study of this problem. As
in \cite{arrow2}, we use as a typical example the periodically
kicked quartic oscillator whose one-step evolution is described
by the Floquet operator ${\cal\hat F}=e^{-\frac{i}{\hbar}{\hat
H}^{(0)}}\, {\hat D}\left(i\frac{g_0}{\sqrt\hbar}\right) =
e^{-i\left(\omega_0 {\hat n}+\hbar {\hat n}^2\right)}\,
e^{i\frac{g_0}{\sqrt{\hbar}}\left({\hat a}+ {\hat
a}^{\dag}\right)}$, where ${\hat a}, {\hat a}^{\dag}$ are the
bosonic ($[{\hat a},{\hat a}^{\dag}]=1$) creation-annihilation
operators and ${\hat n}={\hat a}^{\dag}{\hat a}$ is the
excitation number operator. The driving force is given by
$g(t)=g_0\sum_{\tau} \delta(t-\tau)$. The classical motion of
this oscillator becomes chaotic when the kick strength $g_0$
exceeds unity. We suppose further that each kick is followed by
an instant perturbation $\xi_{\tau}\,\hbar\,{\hat n}
\delta(t-\tau)$ with Gaussian random intensity $\xi_{\tau}$,
($\langle\xi_{\tau}\rangle=0$,
$\langle\xi_{\tau}\xi_{\tau'}\rangle=\sigma^2\delta_{\tau\tau'}$),
which models a persistent external noise. Such a perturbation
gives rise to the phase plane rotation by a random angle
$\xi_{\tau}$ at the time moment $\tau$. For any given
realization (history) $\{\xi\}\equiv\{\xi_1,\xi_2,...,\xi_t\}$
of the noise, the evolution is described by the unitary
operator ${\hat U}(\{\xi\};t)=
\prod_{\tau=1}^{\tau=t}\,\left[e^{-i\xi_{\tau}{\hat
n}}\,{\cal\hat F}\right]$.

We consider below the time evolution of the initially pure
state ${{\hat\rho}(0) =|0\rangle\langle 0|}$, where $|0\rangle$
is the ground eigenstate of the Hamiltonian ${\hat H}^{(0)}$.
The corresponding quantum Wigner function
$W(\alpha^*,\alpha;0)=\frac{2}{\hbar}e^{-\frac{2}{\hbar}|\alpha|^2}$
is isotropic in the phase plane and occupies the phase cell
$\hbar/2$. Few (one in the case of parameters
chosen below) first kicks produce a state of a practically
general form. At a running moment of time $t>0$, the
excitation of the oscillator and the degree of anisotropy of
the Wigner function are characterized by the probability
distributions \cite{arrow2},
\begin{equation}\label{Distr_n}
  w_n(\{\xi\};t)=\langle n|{\hat\rho}(\{\xi\};t)|n\rangle\,,
\end{equation}
and
\begin{equation}\label{Distr_m}
\begin{array}{rl}
 \mathcal{W}_m(\{\xi\};t)&=
  (2-\delta_{m0})\sum_{n=0}^{\infty}
  \Big|\langle n+m\big|{\hat\rho}(\{\xi\};t)\big|n\rangle\Big|^2\\
&=(2-\delta_{m 0})
\sum_{n=0}^{\infty}w_n(\{\xi\};t)w_{n+m}(\{\xi\};t),
\end{array}
\end{equation}
respectively (both normalized to unity). Here
${\hat\rho}(\{\xi\};t)= {\hat U}(\{\xi\};t){\hat\rho}(0){\hat
U}^{\dag}(\{\xi\};t)$. With the noise history being {\it
fixed}, the evolution is unitary so that the state remains {\it
pure} during the whole time of the motion.

{\bf Coarse-grained features of quantum evolution}. The mean
values $\langle n\rangle_{\{\xi\};t}$ and $\langle
|m|\rangle_{\{\xi\};t}$ calculated with the help of the
distributions (\ref{Distr_n},\ref{Distr_m}) characterize
respectively the degree of excitation and the number of
$\theta$-harmonics, i.e. complexity \cite{arrow2} of the
quantum state developed by the time $t$. Our numerical
simulations showed that these values do not depend on the noise
history at a given noise level $\sigma$ (i.e. they are
self-averaging quantities).

\begin{figure}[!h]
 \centering
 \includegraphics[width=75mm,bb=50 190 561 727,keepaspectratio=true]{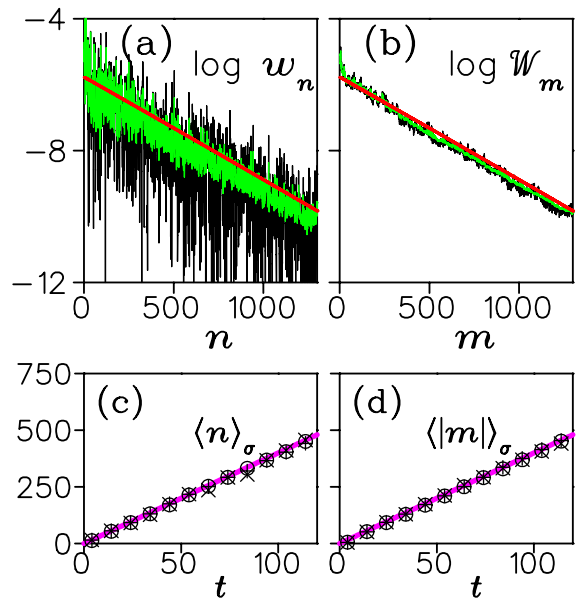}
\caption{(Color online). (a),(b) - Probability
distributions (\ref{Distr_n},\ref{Distr_m}) at the moment
$t=80$ with no noise ($\sigma=0$, black), weak noise
($\sigma=0.001$, green/gray) and in the strong noise limit,
(red/darkgray) In the two latter cases the data have been
averaged over $10^3$ noise realizations. (c),(d)
- Time evolution of the  moments
$\langle n \rangle_\sigma$, $\langle | m | \rangle_\sigma$ (crosses,
pluses and circles for $\sigma=0,0.001$ and $\sigma=\infty$, respectively)
as compared {\bf to} the classical
diffusion law (straight magenta/gray lines).
\label{fig:coarse_grained}}
\end{figure}

As to the distributions (\ref{Distr_n},\ref{Distr_m})
themselves, our detailed numerical data indicate (see upper
panels in Fig.\ref{fig:coarse_grained}; we fix system
parameters as $\omega_0=1, \hbar=1, g_0=2$ throughout the
paper; the Ehrenfest time $t_E<1$ in this case)
that at a given time $t$ they undergo (as functions of $n$ or
$m$, respectively) fluctuations, rather strong in the first
case and much weaker in the second one, around a {\it regular
exponential decay} with identical slopes. These slopes,
contrary to the fluctuations, are not sensitive to the noise
histories. Such universal exponential laws represent the {\it
coarse-grained} distributions \cite{arrow2}
\begin{equation}\label{C_G_Distr_n}
w^{(c.g.)}_n(\sigma;t)=
\frac{1}{\langle n\rangle_{\sigma;t}+1} \left[\frac{\langle
n\rangle_{\sigma;t}}{\langle n\rangle_{\sigma;t}+1}\right]^n,
\end{equation}
and
\begin{equation}\label{C_G_Distr_m}
\mathcal{W}^{(c.g.)}_m(\sigma;t)=
\frac{2-\delta_{m 0}}{2\langle |m|\rangle_{\sigma;t}+1}
\left[\frac{\langle |m|\rangle_{\sigma;t}}
{\langle |m|\rangle_{\sigma;t}+1}\right]^m
\end{equation}
which entirely ignore the fluctuations. The first moments
$\langle n\rangle_{\sigma;t}$ and $\langle
|m|\rangle_{\sigma;t}=\langle n\rangle_{\sigma;t}$ of these
distributions are the free parameters to be used for fitting
the slopes of the actual distributions calculated numerically
and plotted in Fig. \ref{fig:coarse_grained}. Being defined in
such a way, they approximate quite well the corresponding
history-insensitive mean values numerically found with the help
of the {\it exact} distributions
(\ref{C_G_Distr_n},\ref{C_G_Distr_m}). Moreover, the analysis
of our numerical data reveals (see the lower panels in Fig.
\ref{fig:coarse_grained}) that the low moments depend very
weakly on the noise level $\sigma$ as well. {\it The chosen
noise does not influence appreciably not only the number of
harmonics but also the degree of excitation.} Simulations with
truncated excitation bases in the Hilbert space of different
dimensions, $N\gg 1$ up to N=6000, convinced us that the rate
of growth of the quantum mean excitation is practically
indistinguishable from the classical diffusion law $\langle
n\rangle_t=\frac{g_0^2}{\hbar }\,t$ (straight lines in lower
panels in Fig. \ref{fig:coarse_grained}). Being in contrast to
the case of the quantum kicked rotator, this fact agrees with
the absence of localization of the eigenstates of the Floquet
operator ${\cal\hat F}$ \cite{future}. The
increase of the number of angular harmonics $\langle
m\rangle_t=\langle n\rangle_t$ is also linear as distinct from
the classical exponential upgrowth. Furthermore, the
amplitudes of fluctuations of the exact distributions
(\ref{Distr_n},\ref{Distr_m}) are appreciably reduced when the
noise level $\sigma$ is growing.

A natural way of the coarse graining consists in averaging over
realizations $\{\xi\}$ of the noise \cite{Sokolov84}. Indeed,
as it follows from the data presented in Fig.
\ref{fig:coarse_grained}, such an averaging leaves the slopes
practically unchanged but suppresses the fluctuations. The
latter are entirely obliterated at the strong noise limit
$\sigma\rightarrow\infty$. Hence we {\it define} finally the
coarse-grained distributions as
\begin{equation}\label{C_G_Distrs}
\begin{array}{c}
w^{(c.g.)}_n(t)=\overline{w_n(\{\xi\};t)}\big|_{\sigma=\infty}=
w^{(c.g.)}_n(\sigma=\infty;t);\\
\mathcal{W}^{(c.g.)}_m(t)=
 \overline{\mathcal{W}_m(\{\xi\};t)}\big|_{\sigma=\infty}=
 \mathcal{W}^{(c.g.)}_m(\sigma=\infty;t).
\end{array}
\end{equation}
The bar indicates averaging over the noise. Now, the connection
$\langle |m|\rangle_{\infty;t}=\langle n\rangle_{\infty;t}$
directly follows from the exact relation (\ref{Distr_m}) thus
leading to the identity of the slopes. Being plotted, the
coarse-grained distributions
(\ref{C_G_Distr_n},\ref{C_G_Distr_m}) are indistinguishable
from the distributions (\ref{C_G_Distrs}) shown by the
red/darkgray lines in Fig. \ref{fig:coarse_grained}.

{\bf The information entropy.} As mentioned above, the
characteristic number $\langle |m|\rangle_{\sigma;t}$ of
harmonics of the Wigner function at a given moment of time $t$
can serve \cite{Brumer97, arrow2} as a measure of complexity of
the current quantum state. Another possibility is to use
\cite{Peres96} for the same purpose the information entropy.
The latter, by virtue of the probabilistic interpretation, can
be defined as
\begin{eqnarray}\label{Inf_Entropy}
  {\cal I}(t)&=&-\sum_{m=0}^{\infty}\,\mathcal{W}^{(c.g.)}_m(t)\,
  \ln\mathcal{W}^{(c.g.)}_m(t)\nonumber\\
  &\approx& \ln\langle |m|\rangle_{\infty;t}+1+
  \frac{1}{2\langle |m|\rangle_{\infty;t}}+...\,,\,\,(t\gg 1).
\end{eqnarray}
This entropy grows monotonically with time starting from ${\cal
I}(0)=0$. Generally speaking, during the Ehrenfest time
$0<t\lesssim t_E$ the growth is linear with a slope determined
by the classical Lyapunov exponent. Afterwards it slows down to
the logarithmic behavior.

{\bf Sensitivity of quantum evolution to the noise}. As usual,
sensitivity of the motion to the noise could be characterized
by the overlap (``fidelity'')
$F(\{\xi\};t)=\mathrm{Tr}\left[{\hat\rho(t)}\,{\hat\rho(\{\xi\};t)}\right]$
of the states developed by the time $t$ during the evolution
{\it with} or {\it without} influence of the noisy environment.
However, contrary to the low moments of the distributions
(\ref{Distr_n}, \ref{Distr_m}), the quantity defined in such a
way is not a self-averaging one and strongly depends on the
noise history. The appropriate reduced measure is obtained by
averaging over all possible noise realizations
\begin{eqnarray}\label{Fid}
  F(\sigma;t)&=&\overline{F(\{\xi\};t)}\nonumber\\
  &=&\mathrm {Tr}\left[{\hat\rho(t)}\,{\hat\rho^{(av)}(\sigma;t)}\right]=
\overline{\big|\langle 0|{\hat f}(\xi;t)|0\rangle\big|^2}\,.
\end{eqnarray}
The newly defined fidelity depends on the only parameter
$\sigma$ instead of the entire noise history $\{\xi\}$. This
definition brings into consideration the averaged density
matrix
${\hat\rho^{(av)}(\sigma;t)}=\overline{{\hat\rho(\{\xi\};t)}}$
\cite{Zurek03}.  Since
\begin{equation}\label{Av_Den_Matr}
\langle n'|{\hat\rho^{(av)}(\sigma;\tau)}|n\rangle=
e^{-\frac{1}{2}\sigma^2(n'-n)^2}\, \langle n'|{\cal\hat
F}{\hat\rho^{(av)}(\sigma;\tau-1) {\cal\hat F}^{\dag}}|n\rangle,
\end{equation}
the averaging over the noise suppresses the off-diagonal matrix
elements and cuts down the number of harmonics of the
corresponding averaged Wigner function. Notice that the
normalization condition ${\rm Tr}
{\hat\rho^{(av)}(\sigma;\tau)}=1$ holds during the whole
evolution independently of the noise level.

The unitary fidelity operator that appears in the r.h.s. of Eq.
(\ref{Fid}) reads
\begin{eqnarray}\label{Fid_op}
  {\hat f}(\{\xi\};t)&=&{\hat U}^{\dag}(t)\,{\hat U}(\{\xi\};t)\equiv\nonumber\\
  &&{{\cal\hat F}^{\dag}}\,^t\,\prod_{\tau=1}^{\tau=t}
  \left[e^{-i\xi_\tau{\hat n}}\,{\cal\hat F}\right]=
  \prod_{\tau=1}^{\tau=t}
  e^{-i\xi_\tau{\hat n}(\tau)},
\end{eqnarray}
where ${\hat n}(\tau)={{\cal\hat F}^{\dag}}\,^\tau{\hat n}\,
{{\cal\hat F}}^{\,\tau}$ is the Heisenberg evolution of the
operator ${\hat n}$. It is easy to calculate explicitly the
fidelity (\ref{Fid}) for the two limiting cases of weak and
extremely strong noise.

\begin{figure}[h]
 \centering
 \includegraphics[width=75mm,bb=42 369 545 691,keepaspectratio=true]{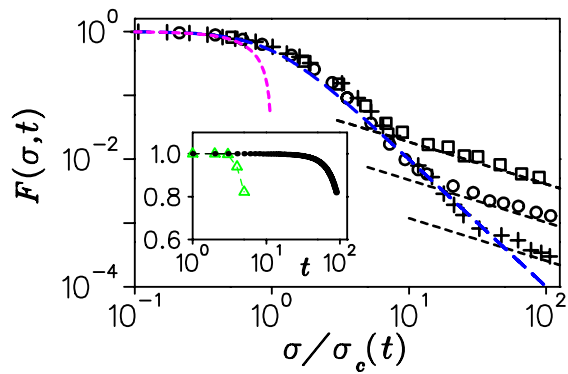}
 % fig2.ps: 503x322 pixel, 72dpi, 17.74x11.36 cm, bb=0 0 503 322
 \caption{(Color online). The scaling properties of the fidelity.
  Crosses, circles and squares correspond respectively to: $\sigma=0.004,
  0.032$ and $0.256$. The magenta/grey and black short-dashed lines
  show the weak and strong noise limits (\ref{Fid_weak}) and
(\ref{Fid_strong2}); the blue/grey long-dashed line corresponds
to the fit (\ref{simple_fit}). Insert: fidelity
decay in the classical Lyapunov ($\hbar=0.01$, green/gray
triangles) and quantum ($\hbar=1$, black circles) regimes; in
the both cases $\sigma=0,0005$. Notice logarithmic scale along
the $t$ axis.} \label{fig:Scaling}
\end{figure}

{\bf Weak noise limit}. Keeping in the last product the noise
intensity $\xi_{\tau}\neq 0$ only in one certain exponential
factor we reduce the problem to that considered in
\cite{arrow2}. Averaging then over $\xi_{\tau}$, expanding up
to the term $\thicksim\sigma^2$ and summing at last over all
the moments $1\leq\tau\leq t$ we obtain
\begin{eqnarray}\label{Fid_weak}
  F(\sigma;t)&=&1-\frac{1}{2}\sigma^2\sum_{\tau=1}^{\tau=t}\sum_{m=1}^{\infty}\,
  m^2\,\mathcal{W}_m(0;\tau)+O(\sigma^4)=\nonumber\\
  && 1-\frac{1}{2}\sigma^2\sum_{\tau=1}^{\tau=t}\langle m^2\rangle_{0;\tau}
  +O(\sigma^4)\,.
\end{eqnarray}
Therefore the fidelity stays close to unity while the noise
level $\sigma$ remains appreciably below the critical value
$\sigma_c(t)=\sqrt{2/\sum_{\tau=1}^{\tau=t}\langle
m^2\rangle_{0;\tau}}\,.$ Insert in
(Fig.\ref{fig:Scaling}) demonstrates moderate decrease of
fidelity during the first 100 kicks ($\hbar=1$). The classical
Lyapunov decay does not show up. This quantum regime is opposed
to the classical fast fall up to the same value in the case
$\hbar=0.01$. Strictly speaking, the mean value $\langle
m^2\rangle_{0;\tau}$ corresponds to the motion with no noise.
However, as it has been already mentioned, the low moments of
the harmonics distribution are practically insensitive to the
noise so that $\langle m^2\rangle_{0;\tau}\approx\langle
m^2\rangle_{\infty;\tau} =2\langle |m|\rangle^2_{\infty;\tau}=
2\langle n\rangle^2_{\infty;\tau}\approx
2\left(\frac{g_0^2}{\hbar}\,\tau\right)^2$. This implies that
$\sigma_c(t)\approx 1/\sqrt{\sum_{\tau=1}^{\tau=t} \langle
|m|\rangle^2_{\infty;\tau}} \propto t^{-3/2}$. Though the
critical value decreases with time faster than in the case of a
single instant perturbation \cite{arrow2}, the decrease is
power-like as before.

{\bf Strong noise limit}. In the opposite case of extremely
strong noise $\sigma\rightarrow\infty$ we insert $t$ times the
completeness condition $\sum_{n=0}^{\infty}|n\rangle \langle
n|=1$ while calculating the matrix element
$f(\xi;t)\equiv\langle 0| {\hat f}(\xi;t)|0\rangle$, then
calculate the average value $\overline {|f(\xi;t)|^2}$ and take
at last into account that $\lim_{\sigma\rightarrow\infty}\,
e^{-\frac{1}{2}\sigma^2 m^2}=\delta_{m 0}$. Finally we arrive
at
\begin{equation}\label{Fid_strong1}
\begin{array}{c}
F_{\infty}(t)=\sum_{n_t}\big|\langle 0|{{\cal\hat
F}^\dag}\, ^{\,t}|n\rangle\big|^2
\sum_{n_{(t-1)}}\big|\langle n_{t}|{\cal\hat F}|n_{(t-1)}\rangle\big|^2\\
...\sum_{n_2}\big|\langle n_3|{\cal\hat F}|n_2\rangle\big|^2
\sum_{n_1}\big|\langle n_2|{\cal\hat F}|n_1\rangle\big|^2
\big|\langle n_1|{\cal\hat F}|0\rangle\big|^2\,.\\
\end{array}
\end{equation}
Only the transition probabilities in the absence of noise are
present, the forward evolution being a chain of non-interfering
successive transitions. Quite opposite, the backward transition
contains the entire set of possible quantum-mechanical paths.

The evolution of the averaged density matrix is of special
interest. Eq. (\ref{Av_Den_Matr}) shows that in the strong
noise limit the density matrix remains diagonal during the
whole time of the motion: $\langle
n'|{\hat\rho^{(av)}(\infty;\tau)} |n\rangle\equiv\langle
n'|{\hat\rho^{(d)}(\tau)}|n\rangle=
\delta_{n'n}\,w^{(d)}_n(\tau)$. The evolution equation
(\ref{Av_Den_Matr}) reduces in this case to
\begin{eqnarray}\label{Mark_chain}
w^{(d)}_n(\tau)&=&\sum_{n'=0}^{\infty}\mathbb{Q}_{nn'}\,w^{(d)}_{n'}(\tau-1),\\
\mathbb{Q}_{nn'}&=&\Big|\langle n|{\hat D}
\left(i\frac{g_0^2}{\sqrt{\hbar}}\right)|n'\rangle\Big|^2>0\,.
\end{eqnarray}
The matrix ${\hat{\mathbb{Q}}}$ is symmetric, positively
definite, and obeys the condition
$\sum_{n'=0}^{\infty}\mathbb{Q}_{n n'}= \langle n|n\rangle=1$.
The equation (\ref{Mark_chain}) describes \cite{Lankaster69} a
homogeneous Markov's chain. Notice that the motion does not
depend in this limit on the properties of the unperturbed
Hamiltonian ${\hat H}^{(0)}.$

It immediately follows now that
\begin{equation}\label{Fid_strong2}
F(\infty;t)=\sum_{n=1}^{\infty}
w_n(0;t)\,w^{(d)}_n(t)\approx \frac{1}{2\langle
n\rangle_{\infty;t}+1}\,.
\end{equation}
We have used here the exponential ansatz (\ref{C_G_Distr_n})
for both $w$ - distributions as well as the practical
independence of the mean excitation number $\langle
n\rangle_{\sigma;t}$ of the noise. This formula is in a good
agreement with our numerical data.

{\bf Moderate noise: scaling property}. Analytical
consideration is not possible in the general case of the
moderate noise level $\sigma\gtrsim\sigma_c(t)$. Generically,
the evolution of the averaged density matrix
${\hat\rho}^{(av)}(\sigma;t)$ is not unitary. This entails
state mixing and suppression of the quantum interference, i.e.
decoherence. Nevertheless, even if this ratio exceeds unity,
the fidelity (\ref{Fid}) continues, as Fig. \ref{fig:Scaling}
clearly demonstrates, to depend only on the ratio
$\sigma/\sigma_c(t)$ up to the time $t_{(dec)}(\sigma)$, when
the full decoherence takes place and the evolution becomes
Markovian.

A simple fit (compare with \cite{arrow2})
\begin{equation}\label{simple_fit}
F_{f}(\sigma;t)=\frac{1}{1+\sigma^2/\sigma_c^2(t)}\,,\quad
 t<t_{(dec)}(\sigma)
\end{equation}
describes our numerical data rather well (Fig.
\ref{fig:Scaling}). With the help of this fit, the decoherence
time is estimated as $t_{(dec)}(\sigma)=
\sqrt{\frac{6\hbar}{D\sigma^2}}$, where $D=g_0^2$ is the
classical diffusion coefficient.

\begin{figure}[h]
 \centering
 \includegraphics[width=75mm,bb=84 383 559 692,keepaspectratio=true]{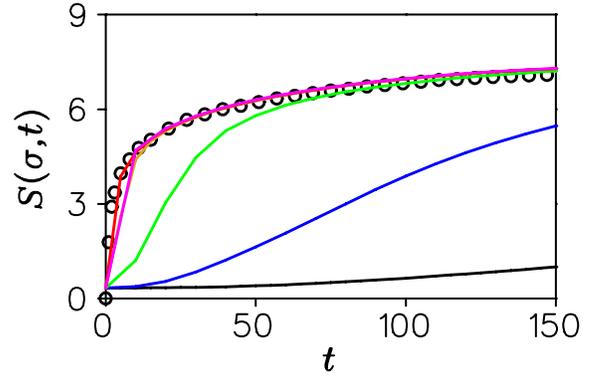}
 % fig3.ps: 475x309 pixel, 72dpi, 16.76x10.90 cm, bb=0 0 475 309
 \caption{(Color online). Von Neumann entropy versus the time.
 From bottom to top (black, blue, green, magenta and red):
 $\sigma= (0.125, 1, 8, 64, 512)\cdot 10^{-3}$. Circles show the
  information Shennon entropy ${\cal I}(t)$,
 see eq. (\ref{Inf_Entropy}).}
 \label{fig:Entropy}
\end{figure}
{\bf Loss of memory on the initial state}. The state mixing
induced by the noise leads to loss of memory about the initial
state. Thereupon the notion of the invariant (independent of
the basis) von Neumann entropy
\begin{equation}\label{Corr_Entropy}
{\cal S}(\sigma;t)= -{\rm
Tr}\left[{\hat\rho}^{(av)}(\sigma;t)\ln{\hat\rho}^{(av)}(\sigma;t)\right]
\end{equation}
becomes relevant. This entropy increases with time monotonically,
Fig. \ref{fig:Entropy}, approaching the function
${\cal I}(t)$ (\ref{Inf_Entropy}) from below when $t\rightarrow
t_{(dec)}$. After the full decoherence takes place, the system
occupies the whole phase volume accessible at the running degree
of excitation $\langle n\rangle_{\sigma;t}\approx\langle n\rangle_{\infty;t}=
\langle |m|\rangle_{\infty;t}$
thus reaching a sort of equilibrium. Henceforth the phase volume expands
"adiabatically": the entropy
${\cal S}(\sigma;t>t_{(dec)})\approx{\cal I}(t)\approx\ln\langle |m|\rangle_{\infty;t}$
remains practically constant
when $\langle n\rangle_{\infty;t}=\langle |m|\rangle_{\infty;t}\gg 1$.
In particular case of the strong noise limit, $(\sigma\rightarrow\infty,\,t_{(dec)}\rightarrow 0)$,
the averaged density matrix is diagonal and the von Neumann entropy equals
\begin{equation}\label{Entropy_S_Noise}
\begin{array}{rl}
 {\cal S}_{\infty}(t)&=-\sum_{n=0}^{\infty}w^{(d)}_n(t)\ln w^{(d)}_n(t)\\
 &\approx\ln\langle n\rangle_{\infty;t}+1+\frac{1}{2\langle n\rangle_{\infty;t}}+...\,
  ,\,\,\,\, (t\gg 1)\,.\\
\end{array}
\end{equation}
We have arrived at a remarkable connection ${\cal S}_{\infty}(t)={\cal I}(t)$.
Similar connection between the information and invariant von Neumann
entropies has been discovered also in the theory of random band matrices
\cite{Sokolov98}.

{\bf Reversibility versus Purity}. Another interesting aspect
of quantum motion under the influence of a noisy background is
the degree of reversibility. This degree can naturally be
measured by the mean overlap (Peres fidelity) of the initial
state ${\hat\rho}(0)$ and the state
${\hat\rho}(0|\{\xi\},\{\xi'\};t)$ formed during forward
evolution for some time $t$ under the influence of a stationary
noise with the level $\sigma$ and a history $\{\xi\}$ and then
backward evolution for the same time under the same noise with
an independent history $\{\xi'\}$:
\begin{equation}\label{Fid_Rev}
\begin{array}{rcl}
  \mathbb{F}(\sigma;t)=
  &\overline{\mathrm{Tr}\left[{\hat\rho}(0)\,{\hat\rho}(0|\{\xi\},\{\xi'\};t)\right]}=&\\
  &\overline{\mathrm{Tr}\left[{\hat\rho}(\{\xi\};t)\,{\hat\rho}(\{\xi'\};t)\right]}=&\\
  &{\rm Tr}\left[{\hat\rho}^{(av)}(\sigma;t)\,{\hat\rho}^{(av)}(\sigma;t)\right]&
   \equiv\mathbb{P}(\sigma;t)\,.\\
\end{array}
\end{equation}
The quantity $\mathbb{P}(\sigma;t)$ is referred for as {\it
purity} \cite{Zurek03}. It can be expressed in terms of the
mean number of harmonics $\langle |{\mathfrak
m}|\rangle_{\sigma;t}$ of the {\it averaged} Wigner function.
The corresponding probability distribution is related to the
averaged density matrix ${\hat\rho}^{(av)}(\sigma;t)$ as
\begin{equation}\label{Distr_mfrak_m}
\mathbb{W}_m(\sigma;t)=(2-\delta_{m 0})
\frac{\sum_{n=0}^{\infty} \Big|\langle
n+m\big|{\hat\rho}^{(av)}(\sigma;t)\big|n\rangle\Big|^2}{\mathbb{P}(\sigma;t)}\,.
\end{equation}
(to be confronted with eq.({\ref{Distr_m}})). This means in particular that,
in the exponential approximation for mixed states \cite{arrow2},
\begin{eqnarray}\label{Pur_vs_W_0}
\mathbb{P}(\sigma;t)&=&\frac{1}{\mathbb{W}_0(\sigma;t)}\sum_{n=0}^{\infty}
\left[w^{(av)}_n(\sigma;t)\right]^2\approx\nonumber\\
&&\frac{2\langle |{\mathfrak m}|\rangle_{\sigma;t}+1}
 {2\langle n\rangle_{\sigma;t}+1}\xrightarrow{\sigma\rightarrow\infty}
 \frac{1}{2\langle n \rangle_{\infty;t}+1}\,.
\end{eqnarray}
Contrary to $\langle |m|\rangle_{\sigma;t}$, the mean value
$\langle |{\mathfrak m}|\rangle_{\sigma;t}$ strongly depends on
the noise level and vanishes rapidly when $\sigma$ grows.
Comparison with Eq. (\ref{Fid_strong2}) shows that, as in
\cite{arrow2}, the degrees of reversibility and sensitivity to
external perturbations are directly connected,
$\mathbb{F}(\infty;t)=F(\infty;t)$. Notice also the relation
${\cal S}_{\infty}(t)\approx -\ln
\mathbb{P}_{\infty}(t),\,(t\gg 1)$ that follows from Eqs.
(\ref{Entropy_S_Noise},\ref{Pur_vs_W_0}).

{\bf Summary}. The main goal of this Letter was to investigate
in detail the the dynamics of a classically chaotic
quantum system with few (one in our illustrative model) degrees
of freedom affected by a persistent external noise under the
condition that the Ehrenfest time interval is so short that
the classical-like exponential instability does not show up.
We have shown first that the noise weakly
influences the complexity of the quantum state, which we
characterize by the mean number $\langle |m|\rangle_{\sigma;t}$
of $\theta$-harmonics of the Wigner function. This number
almost does not depend on the noise realization (self-averaging
property) as well as on the level of the noise. At the same
time the noise efficiently washes off fluctuations of the
corresponding probability distribution thereby displaying the
universal regular exponential decay of the coarse-grained
distribution that describes the features of the motion
independent of the realization of the noise.

The Peres fidelity that specifies a quantitative measure of
sensitivity of the motion to the noise utilizes the density
matrix ${\hat\rho}^{(av)} (\sigma;t)$ {\it averaged over the
noise}. The sensitivity remains weak until the noise level
$\sigma$ exceeds some critical value. We have proved that
with the assumptions indicated above
the decrease of this critical value is power-like,
$\sigma_c(t)\approx 1/\sqrt{\sum_{\tau=1}^{\tau=t} \langle
|m|\rangle^2_{\infty;\tau}} \propto t^{-3/2}$.

A scaling behavior has been discovered: the Peres
fidelity depends only on the ratio $\sigma^2/\sigma_c^2(t)$ in
a wide interval of the noise level up to some value
$\sigma_d(t)$ that is considerably larger than the critical
value $\sigma_c(t)$. The scaling is destroyed only under
influence of even a stronger noise $\sigma>\sigma_d(t)$. The
evolution becomes Markovian in this case. This implies that
decoherence takes place at the time $t_{(dec)}(\sigma)\thicksim\sqrt{\frac{\hbar}{D\sigma^2}}$.
The information entropy and the invariant von Neumann entropy
coincide, ${\cal S}_{\sigma}(t)\Rightarrow{\cal I}(t)$, when
$t\gtrsim t_{(dec)}(\sigma)$. They are identical in the limit
$\sigma\rightarrow\infty$. We have also noticed that the
reversibility of the motion influenced by a persistent noise is
measured by the purity of the state at the moment of time
reversal, $\mathbb{F}(\sigma;t)=\mathbb{P}(\sigma;t)$.

\acknowledgments
 We would like to thank F. Borgonovi,
A. Buchleitner and G. Mantica for their interest to this work
and useful remarks. We acknowledge financial support by RFBR
(grant № 09-02-01443) and by the RAS Joint scientific program
"Nonlinear dynamics and Solitons".

\end{document}